\documentclass[aip,
jcp,
amsmath,
amssymb,
reprint, 
citeautoscript,
]{revtex4-2}

\usepackage{graphicx}
\usepackage{txfonts}

\begin{document}

\preprint{}

\title[
Recursive Algorithm to the Centroid of Free Area for Inherent Structure and Hopping Motion in Deeply Supercooled Binary Hard Disk Systems
]{
Recursive Algorithm to the Centroid of Free Area for Inherent Structure and Hopping Motion in Deeply Supercooled Binary Hard Disk Systems
}

\author{Daigo Mugita}
\affiliation{Graduate School of Engineering, Nagoya Institute of Technology,
Nagoya, 466-8555, Japan}

\author{Kazuyoshi Souno}
\affiliation{Graduate School of Engineering, Nagoya Institute of Technology,
Nagoya, 466-8555, Japan}

\author{Masaharu Isobe}
 \email{isobe@nitech.ac.jp}
\affiliation{Graduate School of Engineering, Nagoya Institute of Technology,
Nagoya, 466-8555, Japan}

\date{\today}

\begin{abstract}
Inherent structures, derived by eliminating thermal fluctuations from complex trajectories, illuminate fundamental mechanisms underlying structural relaxation and dynamic heterogeneity in dense glassy systems. However, determining these structures in hard disk/sphere systems presents unique challenges due to the discontinuous nature of inter-particle potentials and resultant flat potential energy landscapes. To address this limitation, we introduce the Recursive Centroid of Free Area algorithm (ReCFA), a novel approach inspired by a steepest descent method, which computes inherent structure configurations in hard disk systems.
We conducted comparative analyses between ReCFA, similar methods, and a conventional time-coarse-graining technique, focusing on string-like hopping motions in supercompressed binary hard disks that emulate supercooled liquid behavior. ReCFA demonstrated notable advantages in capturing entropic contributions. The configurations derived through ReCFA exhibited physically reasonable particle displacements analogous to inherent structures in soft particle systems, effectively identifying hopping motions between metastable basins in jammed states.
This ReCFA-based analysis enhances our understanding of relaxation dynamics in highly compressed glassy systems, offering a robust analytical tool for investigating both dynamic and structural characteristics across hard and soft particle systems.
\end{abstract}


\maketitle

\section{\label{sec:1} Introduction}
Understanding the microscopic mechanisms underlying the dramatic slowdown in dynamics of deeply supercooled liquids in glass-forming materials remains a fundamental challenge in condensed matter physics~\cite{chandler_2010, berthier_2011, biroli_2013, royall_2015, scalliet_2022}. The physics of glassy systems, including their thermodynamics, has been extensively described through the topography of the potential energy landscape, which characterizes the potential energy as a function of the system's coordinates~\cite{goldstein_1969, cavagna_1998, auffinger_2013, ros_2019}. Stillinger and Weber proposed a concrete analytical framework for the potential energy landscape using ``inherent structures,'' which comprise collections of nearest minima within each basin, representing purely structural properties~\cite{stillinger_1982, stillinger_1984, stillinger_1995}. From a statistical mechanical perspective, this corresponds to a tiling of configuration space, represented by the ensemble of inherent structures.
While the approach of understanding glassy system dynamics and thermodynamics through inherent structures has been subject to debate regarding its validity~\cite{berthier_2003, dyre_2006}, it has generated significant interest~\cite{heuer_2008, baity_2021}. Studies employing inherent structures, which provide clear representations of structural properties by eliminating thermal noise, have yielded valuable insights across diverse fields, including sound propagation~\cite{gelin_2016}, protein dynamics~\cite{rao_2010}, and artificial neural networks~\cite{bansal_2018}.

In soft sphere systems, the inherent structure representing the tile to which a given configuration belongs is typically obtained through steepest descent~\cite{nocedal_2006}, involving an instantaneous quenching process of recursively displacing particles along the potential energy gradient. 
Since the free energy change
$\Delta F$ satisfies
$\Delta F = \Delta U$ 
as
$T \to 0$, such recursive operations can 
be used to
derive the state obtained 
through
quenching~\cite{donev_2007a}, where $\Delta U$
represents
the change in potential energy and $T$ 
denotes
the temperature. Recent detailed investigations of this relaxation dynamics have revealed non-trivial power-law slow dynamics across multiple systems~\cite{chacko_2019, gonzalez_2020, folena_2020, charbonneau_2021, folena_2021, nishikawa_2022, stanifer_2022, manacorda_2022}. Beyond steepest descent, more efficient optimization methods, such as conjugate gradient (CG)~\cite{nocedal_2006} and fast inertial relaxation engine (FIRE)~\cite{bitzek_2006}, are frequently employed to enhance computational efficiency in relaxation processes.

In hard sphere systems, the potential energy landscape between particles is flat, and intrinsic thermodynamic properties are temperature-independent. For instance, low-temperature systems exhibit behavior consistent with high-temperature systems in slow motion. Therefore, at fixed positive temperature $T$, the free energy change can be expressed as $\Delta F=-T\Delta S$ using the entropy change $\Delta S$. In hard sphere systems, considering the limit $T \to 0$ as in soft sphere systems yields states that clearly do not qualify as ``inherent,'' since motion ceases immediately at positions just prior to quenching. Instead, inherent structures in hard sphere systems can be obtained by increasing the packing fraction $\nu$ and considering the infinite pressure limit $P \to \infty$. In this scenario, particle diameters expand uniformly, with contacted particles being pushed against each other until reaching a jammed state~\cite{stillinger_1964}. This approach yields inherent structures because, in soft sphere systems, such structures correspond to the zero-pressure limit where internal stresses vanish, and these configurations are known to correspond to collectively jammed states~\cite{stillinger_1985, hern_2003, donev_2004, donev_2007a, torquato_2001}. More detailed correspondences with soft sphere systems are presented in Table1 of Ref.~\onlinecite{donev_2007a}.

The extensive history of jammed packing studies, originating from theories of amorphous solid glasses and supercooled liquids~\cite{stillinger_1964, scott_1969, berryman_1983, jodrey_1985, meakin_1993, torquato_2001, torquato_2010}, has produced numerous methods for jamming hard spheres~\cite{tory_1968, adams_1972, kansal_2000, woodcock_1976, jodrey_1981, lubachevsky_1990, lubachevsky_1991, zinchenko_1994, speedy_1998, artiaco_2022}. However, most of these methods are unsuitable for deriving inherent structures in hard sphere systems. Many conventional approaches, including algorithms that sequentially add particles to existing configurations~\cite{tory_1968, adams_1972, kansal_2000} or those based on dynamic processes~\cite{woodcock_1976, jodrey_1981, lubachevsky_1990, lubachevsky_1991}, generate jammed structures without considering specific instantaneous configurations, thereby failing to accurately represent inherent structures. Among these methods, Speedy's algorithm~\cite{speedy_1998} is notable for generating inherent structures by recursively moving each particle in a $d$-dimensional system toward the center of the cage formed by its $d+1$ nearest neighbors, while the particles expand.

A significant application of inherent structures lies in analyzing string-like hopping motions in supercooled liquids. These motions involve particle displacements on the order of their diameter, occurring in chain-like sequences across multiple particles. This phenomenon is considered a fundamental structural relaxation mechanism underlying the dramatic viscosity increase with decreasing temperature and has been widely observed in molecular dynamics simulations and colloidal experiments across various glass-forming systems~\cite{miyagawa_1988, keys_2011, speck_2012, helfferich_2014, isobe_2016b, schoenholz_2016, lam_2017, lam_2018, yip_2020}. Recent studies have demonstrated that back-and-forth string-like hopping motions, characterized by particles returning to their original positions after hopping, become predominant as density increases and temperature decreases~\cite{lam_2017, yip_2020}.

Hopping motions are associated with significant dynamic changes in inherent structure. In soft sphere systems, hopping analyses typically utilize particle trajectories in the inherent structure to avoid capturing anharmonic vibrations within basins~\cite{speck_2012, schoenholz_2016}. In hard sphere systems, trajectories processed by the conventional time-coarse-graining (TCG) method~\cite{belch_1981, hirata_1981} have traditionally been employed~\cite{isobe_2016b, yip_2020}. While TCG-processed trajectories show some agreement with inherent structure trajectories~\cite{keys_2011}, particularly when large-displacement transitions between basins do not occur, TCG remains an approximation method requiring time resolution to average molecular dynamics (MD) trajectories as an adjustable parameter. Consequently, TCG trajectories cannot be considered inherent structures of configurations at ``instantaneous'' moments.

In this study, we aim to achieve more precise hopping motion analysis compared to conventional TCG. We propose a novel method for deriving nearly jammed packings and inherent structures in two-dimensional (2D) hard sphere systems (i.e., hard disk systems), termed ``Recursion algorithm to the Centroid of Free Area (ReCFA),'' inspired by the steepest descent method. Rather than moving particles along potential energy gradients, this method recursively moves particles toward the centroid of the free volume (free area in 2D). The free volume is defined as the space available to a tagged particle's center when surrounding particles are fixed and is efficiently calculated using the 
method of categorizing neighbors for enclosing the local
free area (NELF-A)
developed by the authors~\cite{mugita_2024}. In addition to movements toward the centroid of the free area (CFA), we explored movements toward the ``inner center'' (analogous to the incenter) of free area (IFA) and Speedy's cage center (CC). While these algorithms successfully created nearly jammed packings, they were unable to eliminate thermal fluctuations from particle trajectories.

This paper is structured as follows. Section~\ref{sec:2-0} details the recursive algorithms for obtaining nearly jammed packings and defines the three centers. Section~\ref{sec:3-0} investigates their relaxation dynamics in terms of recursive procedures and the properties of their generated states, both in monodisperse system solid phases and bidisperse system glassy states. Section~\ref{sec:4-0} presents hopping analyses using cage-relative displacements from both ReCFA-derived coordinates and TCG-processed coordinates, comparing results and discussing observed differences. Concluding remarks are provided in Sec.~\ref{sec:5-0}.

\section{Method}
\label{sec:2-0}
We implemented three recursive methods to generate nearly jammed packing, analogous to the steepest descent method, to investigate hopping motion analysis. These methods differ solely in the direction of particle movement. In each algorithm iteration, particles move toward one of three centers: the ``inner center'' of the free area (IFA), the Speedy's cage center (CC), or the centroid of the free area (CFA). Below, we describe the recursive algorithm and these three centers in detail.

\subsection{Recursive algorithm}
\begin{figure}
\begin{center}
\includegraphics[scale=0.13]{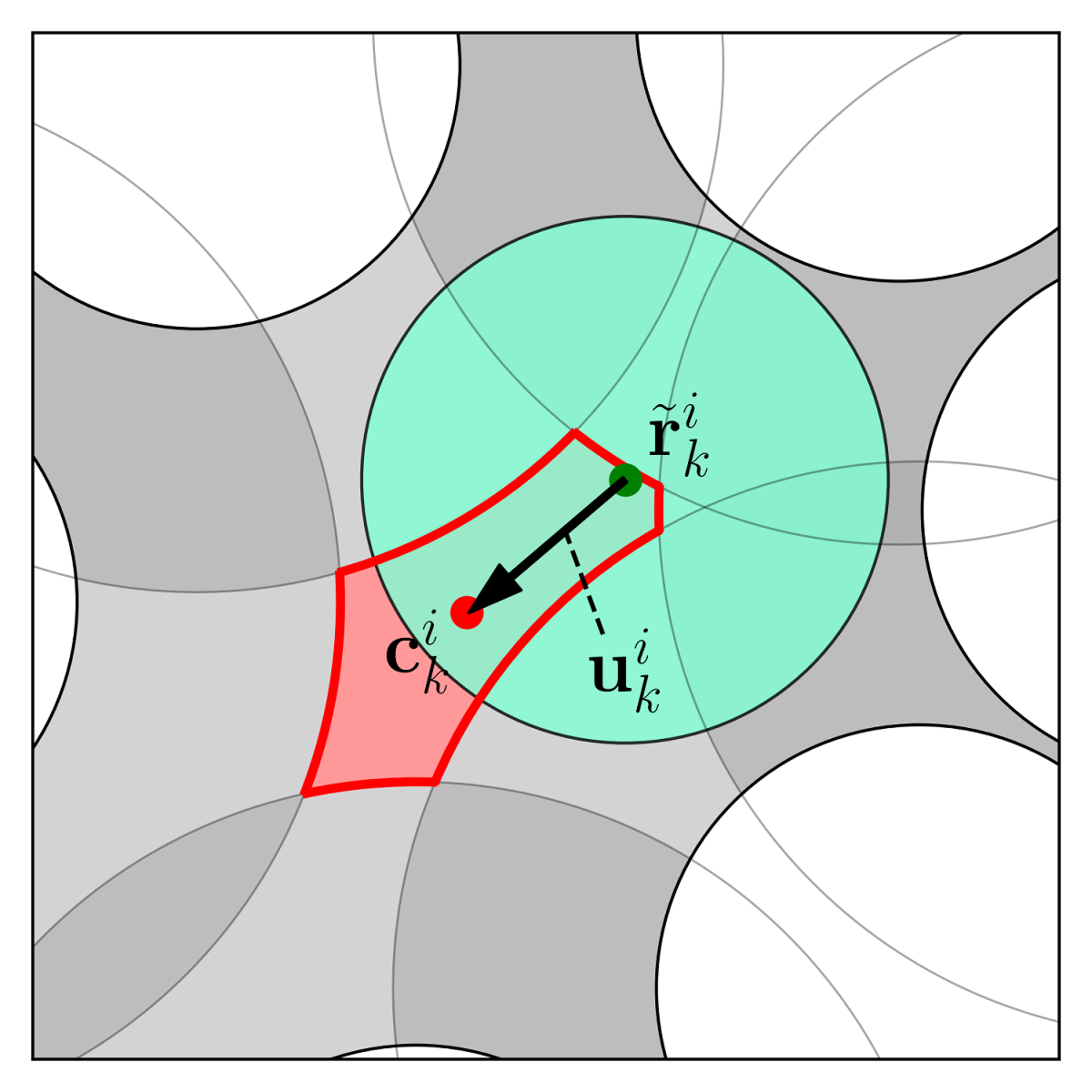}
\caption{ 
The displacement direction vector $\mathbf{u}^i_{k}$ connects the center of the tagged particle $\tilde{\mathbf{r}}^i_{k}$ (green point) to its respective ``center'' $\mathbf{c}^i_{k}$ (red point). This figure illustrates the case of CFA.}
\label{fig:gc}
\end{center}
\end{figure}
The details of the recursive algorithm are explained as follows. 
Here, $\mathbf{r}^i$ represents the initial position of particle $i$, $\tilde{\mathbf{r}}_k^i$ denotes the position at the $k$-th iteration ($k = 0,\  1,\  2,\  \dots$), and $\mathbf{c}_k^i$ represents the position of respective ``center''.
\begin{itemize}
\item[1.] (Except for CC) Calculate the current free area for all particles in the system using NELF-A~\cite{mugita_2024}.
\item[2.] Draw the displacement direction vectors $\mathbf{u}^i_{k}=\mathbf{c}^i_{k}-\tilde{\mathbf{r}}^i_{k}$ connecting the points from the center of individual particle to its respective ``center'' for all particles in the system (see Fig.~\ref{fig:gc} for the case of CFA).
\item[3.] Displace the current position vectors $\tilde{\mathbf{r}}^i_{k}$ for each particle $i$ to new ones $\tilde{\mathbf{r}}^i_{k+1}$ by adding the correction term $\alpha\mathbf{u}^i_{k}$, where $\alpha$ is a scaling factor, i.e.,
\begin{equation}
\tilde{\mathbf{r}}^i_{k+1}=\tilde{\mathbf{r}}^i_k+\alpha\mathbf{u}^i_k \ (\tilde{\mathbf{r}}^i_0=\mathbf{r}^i \text{ at } k=0).
\label{eqn:rec}
\end{equation}
\item[4.] Eliminate the minute overlap by applying a translational displacement to the particle, moving it away from the adjacent particles with which it intersects.
\item[5.] Repeat steps 1 -- 4 for $K$ times until the particle positions converge, obtaining the nearly jammed state $\mathbf{r}^i_{\mathrm{jam}}(=\tilde{\mathbf{r}}^i_K)$. Convergence is quantified under the condition that
\begin{equation}
\langle u\rangle=\frac{1}{N}\sum_{i=1}^N\left|\mathbf{u}^i_k-\langle \mathbf{u}_k \rangle \right|,
\label{eqn:u}
\end{equation}
\begin{equation}
\langle u\rangle^\ast=\langle u\rangle/(2\sigma)\le\xi,
\label{eqn:condition}
\end{equation}
where $N$ is the number of particles in a system, $\sigma$ is the particle radius, and $\xi$ is the threshold for detecting convergence. Angle brackets denote the average for all particles in the system.
\end{itemize}
The reason for subtracting $\langle \mathbf{u}_k \rangle$ in Eq.~(\ref{eqn:u}) is that in the recursive algorithm using IFA and CFA in a binary system, a unidirectional flow was sometimes observed.

A similar concept to this was proposed for the Lloyd algorithm~\cite{lloyd_1982}, which can generate centroidal Voronoi tessellations~\cite{du_1999}, and has recently been used in research on the Quantizer problem~\cite{klatt_2019}.

\subsection{Definition of the centers}
\begin{figure}
\begin{center}
\includegraphics[scale=0.055]{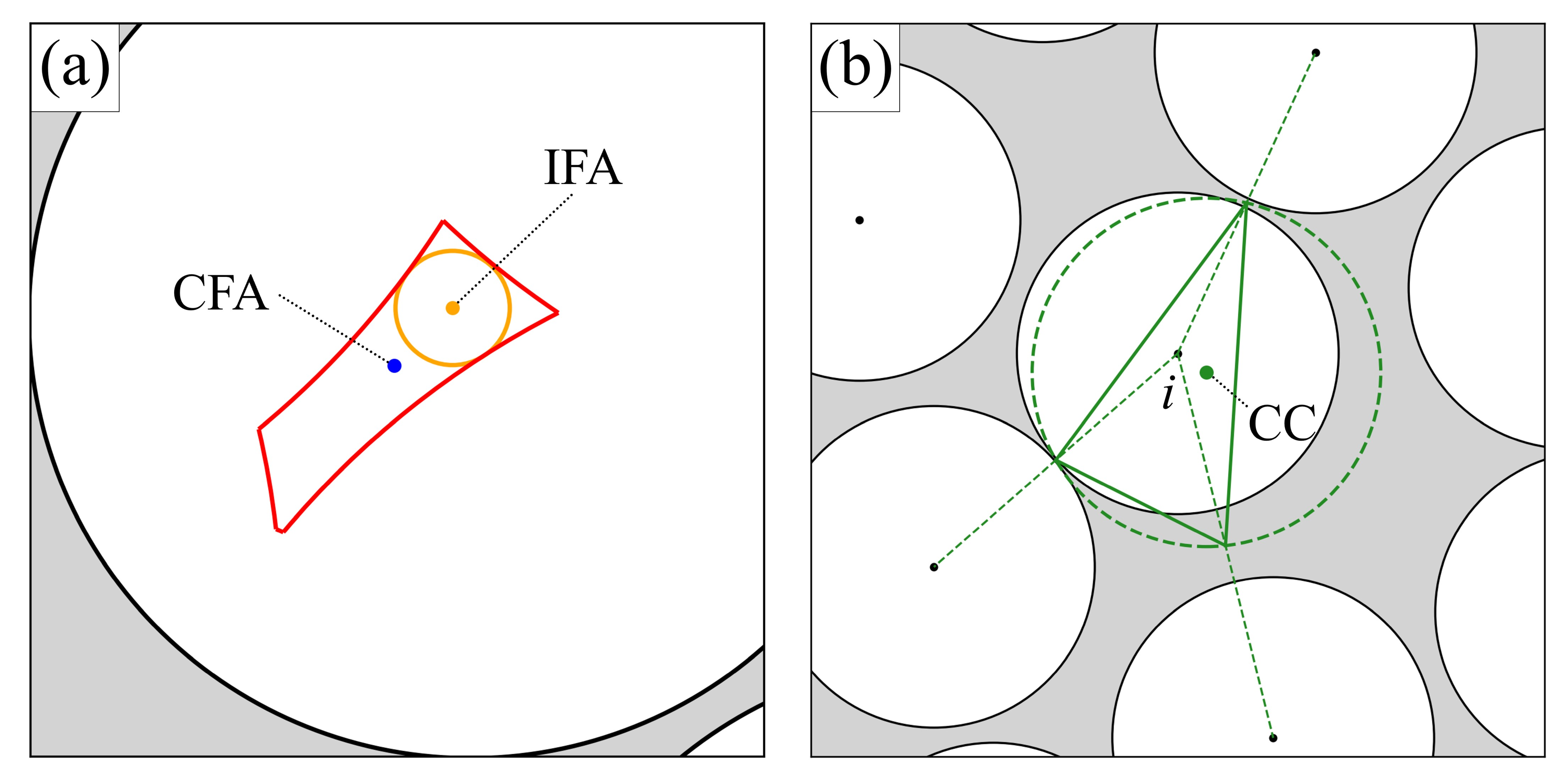}
\caption{(a) The IFA (orange point) and CFA (blue point) of the free area (delineated by red curves). (b) The cage (green triangle) of particle $i$ and its center, i.e., the CC (green point) in a monodisperse system.}
\label{fig:def}
\end{center}
\end{figure}
In Step 2 of the recursive algorithm, three different ``centers''---IFA, CC, and CFA---were employed. Their explicit definitions and numerical methods are described below.
First, we discuss the ``inner center'' of the free area (IFA). While incenter, defined as the center of the inscribed circle, is generally applicable only to triangles, we define the IFA as the center of the largest circle that can be inscribed within the free area, as illustrated in Fig.~\ref{fig:def}(a). The position of this center was determined using the Monte Carlo method.
Next, we consider the cage center (CC), as defined by Speedy~\cite{speedy_1998}. The definition and numerical method for determining the cage and its center in a 2D system are as follows: For each particle $i$ and its neighbors $j$, the point $q_{ij}$ is found by internally dividing the vector from the center of $i$ to the center of $j$ in the ratio of their radii. In a monodisperse system, $q_{ij}$ is the midpoint of the line connecting the centers of $i$ and $j$. The triangle is then constructed using the set of $q_{ij}$ points, $q_{i{j}}\ (j=1,\ 2,\ 3)$, as vertices. Only triangles are considered where the center of $i$ lies within, and among these, the triangle with the smallest circumscribed circle is defined as the cage of $i$. The CC is defined as the center of this circumscribed circle (see Fig.~\ref{fig:def}(b)).
Finally, we describe the method for calculating the centroid of the free area (CFA). In general, the centroid position vector $\mathbf{g}$ in arbitrary closed areas can be decomposed into $n$ fragmented pieces: $\mathbf{g}=\sum_{l=1}^n\mathbf{g}^lA^l/\sum_{l=1}^nA^l$, where $\mathbf{g}^l$ and $A^l$ are the centroid position vector and the area of the tagged fragment piece $l$, respectively. For the free volume (area) $v_{\mathrm{f}}^i$ of particle $i$, as illustrated in Fig.~11 of Ref.~\onlinecite{mugita_2024}, by considering the large triangle constructed from the free area, multiple sectors, and multiple small triangles, the centroid position vector of free area $\mathbf{g}_{\mathrm{f}}^i$ for each particle $i$ can be calculated as follows:
\begin{equation}
\mathbf{g}_{\mathrm{f}}^i=\frac{1}{v_{\mathrm{f}}^i}\left\lbrack\mathbf{g}_{\mathrm{T}}v_{\mathrm{T}}-\sum_{l=1}^{N_{\mathrm{f}}}(\mathbf{g}_{\mathrm{t}}^lv_{\mathrm{t}}^l+\mathbf{g}_{\mathrm{s}}^lv_{\mathrm{s}}^l)\right\rbrack,
\label{eqn:gc}
\end{equation}
where $N_{\mathrm{f}}$ is the number of particles constructing the free area around $i$. $v_{\mathrm{T}}$ is the area of the large triangle, and $v_{\mathrm{t}}^l$ and $v_{\mathrm{s}}^l$ are the areas of the small triangles and the sectors tagged with $l$, respectively. Similarly, $\mathbf{g}_{\mathrm{T}}$ is the centroid vector of the large triangle, and $\mathbf{g}_{\mathrm{t}}^l$ and $\mathbf{g}_{\mathrm{s}}^l$ are the centroid vectors of the small triangles and sectors tagged with $l$, respectively.

\subsection{Characterization of center types}
\begin{figure*}
\centering
\includegraphics[scale=0.10]{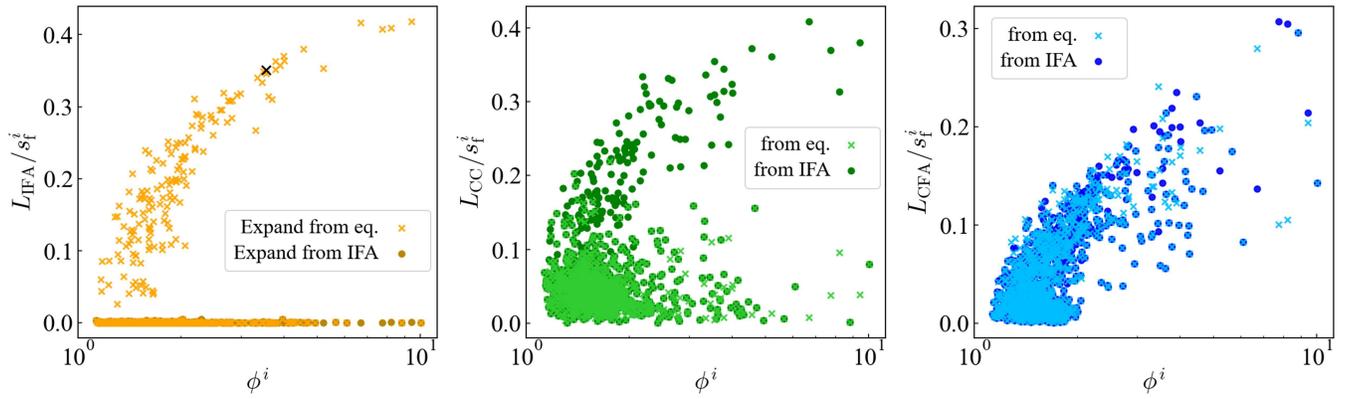}
\caption{Distances $L_{\mathrm{IFA}}$, $L_{\mathrm{CC}}$, and $L_{\mathrm{CFA}}$ from IFA, CC, and CFA to the expanded position versus circularity $\phi^i$, normalized by free surface area $s_{\mathrm{f}}^i$. Crosses denote expansion from equilibrium, while solid circles indicate expansion from IFA.
}
\label{fig:expand}
\end{figure*}
In steepest descent, the potential energy gradient is calculated for a target particle $i$, with other particles fixed, to reach a local minimum. We examine the properties of three centers (IFA, CC, and CFA) in a monodisperse solid phase $(N, \nu) = (1024, 0.740)$, where $\nu$ is the packing fraction. The details of the system setup are given in Sec.~\ref{sec:3-0}.

\begin{figure}
\begin{center}
\includegraphics[scale=0.075]{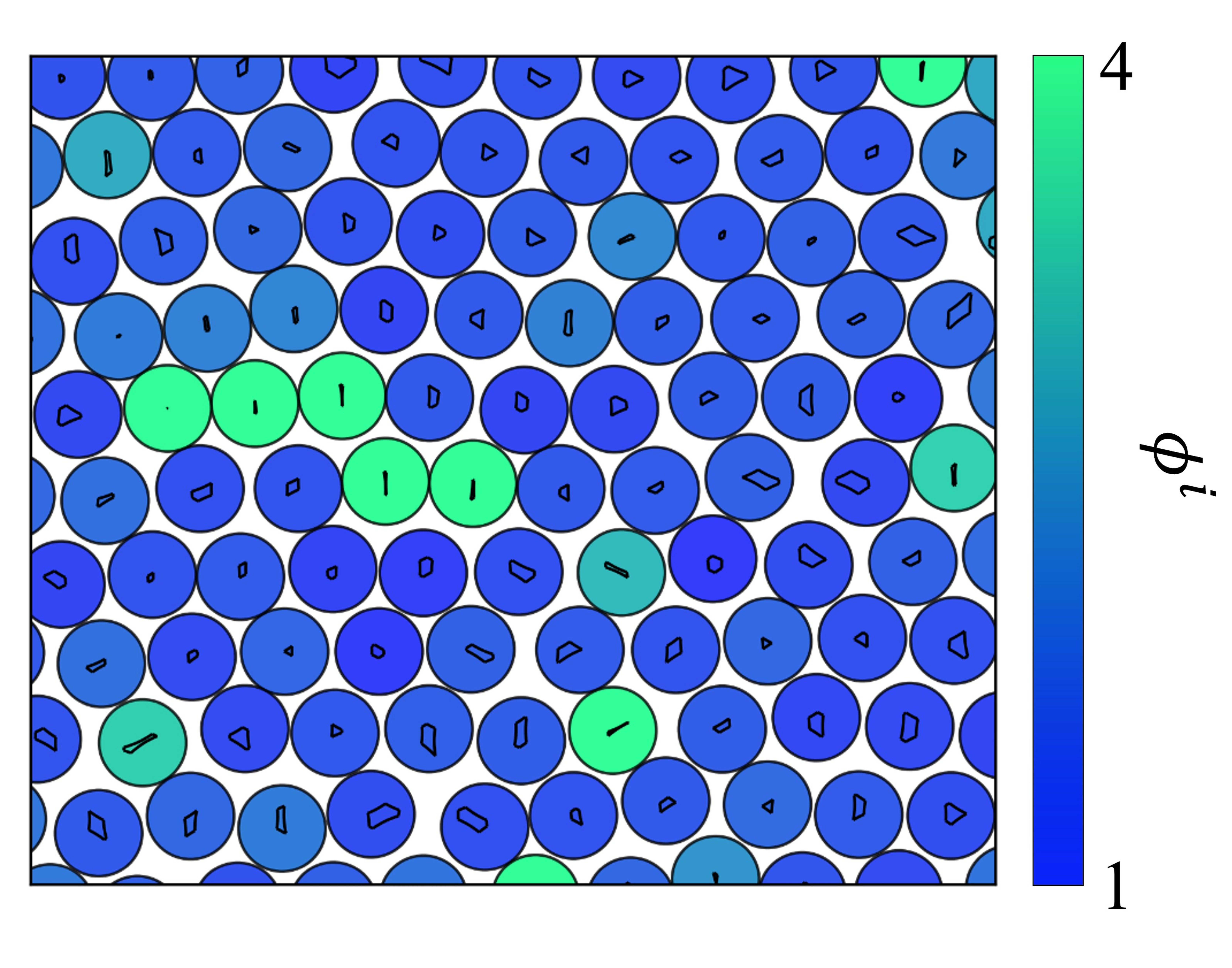}
\caption{
Free area shapes and circularity $\phi^i$ ($\geq 1$) in a monodisperse system $(N, \nu)=(1024, 0.740)$. $\phi^i$ increases with the free area anisotropy.
}
\label{fig:phi}
\end{center}
\end{figure}

To create nearly jammed packings, we compare the expanded position of the particle $i$ with the three centers. The expansion process involves: (i) expansion of only the particle diameter $i$, (ii) resolving overlaps by translational motion opposite to overlapping neighbors, and (iii) for multiple overlaps, moving in the direction of summed opposite vectors. 
The ``expanded position'' is defined where the particle $i$ contacts at least 3 particles $(=d+1)$, with no contacts in the same semicircle. The distances $L_{\mathrm{IFA}}$, $L_{\mathrm{CC}}$, and $L_{\mathrm{CFA}}$ from this position to the respective centers are calculated for all $1024$ particles and plotted against the circularity $\phi^i$ (Fig.~\ref{fig:expand}). 
Circularity $\phi^i$ is defined as:
\begin{equation}
\phi^i = \frac{{s_{\mathrm{f}}^i}^2}{4\pi v_{\mathrm{f}}^i},
\end{equation}
where $v_{\mathrm{f}}^i$ and $s_{\mathrm{f}}^i$ are free volume and surface area. $\phi^i = 1$ for a circle, increasing with anisotropy (Fig.~\ref{fig:phi}).

\begin{figure}
\begin{center}
\includegraphics[scale=0.085]{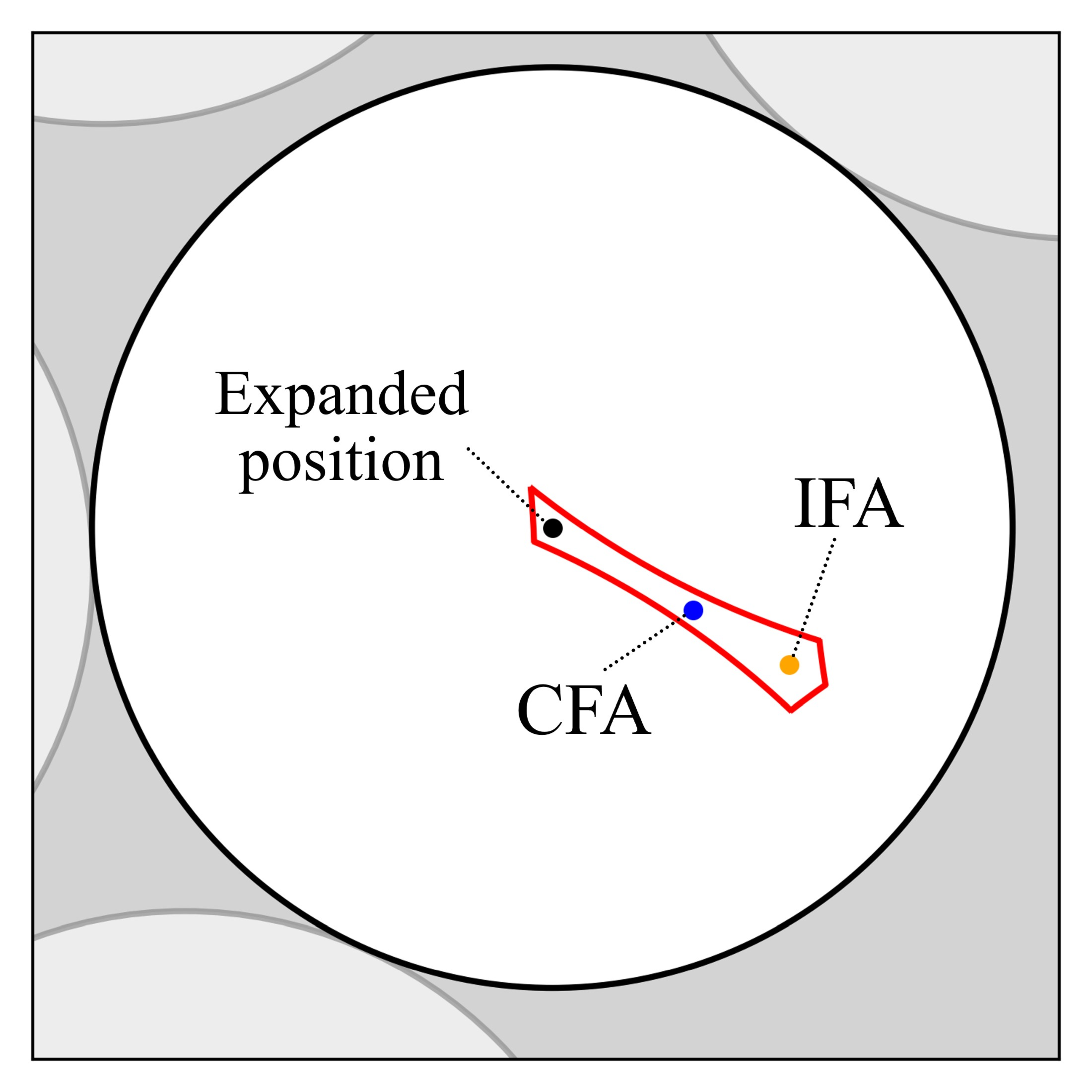}
\caption{
Double convergence state for the target particle (black cross in Fig.~\ref{fig:expand}, $(\phi^i, L_{\mathrm{IFA}}/s_{\mathrm{f}}^i) = (3.55, 0.35)$). The particle outlined in black represents the target particle in its expanded state. The black point indicates the expanded position from equilibrium, while orange and blue points denote IFA and CFA, respectively.
}
\label{fig:double}
\end{center}
\end{figure}

For IFA, the particle scatter plots show two domains: $L_{\mathrm{IFA}} \simeq 0$ and $L_{\mathrm{IFA}} \neq 0$, the latter correlating with $\phi^i$. Fig.~\ref{fig:double} illustrates ``double convergence states'' for $L_{\mathrm{IFA}} \neq 0$, explaining non-coincidence of expanded position and IFA. The larger $\phi^i$ correlates with the elongated free areas in these states. 
Expansion from IFA always results in coincidence (solid filled circles in Fig.~\ref{fig:expand}) with larger expansion radius, suggesting IFA allows maximum expansion with fixed surroundings. 
IFA may correspond to the deepest accessible potential energy minimum in soft particle systems, though direct comparisons require further discussion.

For CC, $L_{\mathrm{CC}}$ from equilibrium is smaller than from IFA, reversing the trends for crosses and solid circles compared to IFA. Unlike $L_{\mathrm{IFA}}$, the $L_{\mathrm{CC}}$ crosses are widely distributed, preventing an approximation to zero. Thus, CC represents an approximate nearest convergent position, potentially corresponding to the nearest potential energy minimum in soft particle systems.

For CFA, there were no significant differences in $L_{\mathrm{CFA}}$ between the expansion from equilibrium and from IFA, explained by the position of CFA between convergent points in double convergence states (see Fig.~\ref{fig:double}). $L_{\mathrm{CFA}}$ shows weak correlation with $\phi^i$. High $\phi^i$ does not always indicate double convergence (e.g., diamond-shaped free area). However, in monodisperse solid phases (Fig.~\ref{fig:phi}), particles with large $\phi^i$ are often constrained by opposing particles along the crystal axis, creating anisotropic free areas. This leads to double minimum states for large $\phi^i$ particles, correlating $L_{\mathrm{CFA}}$ and $\phi^i$.

\begin{figure}
\begin{center}
\includegraphics[scale=0.2]{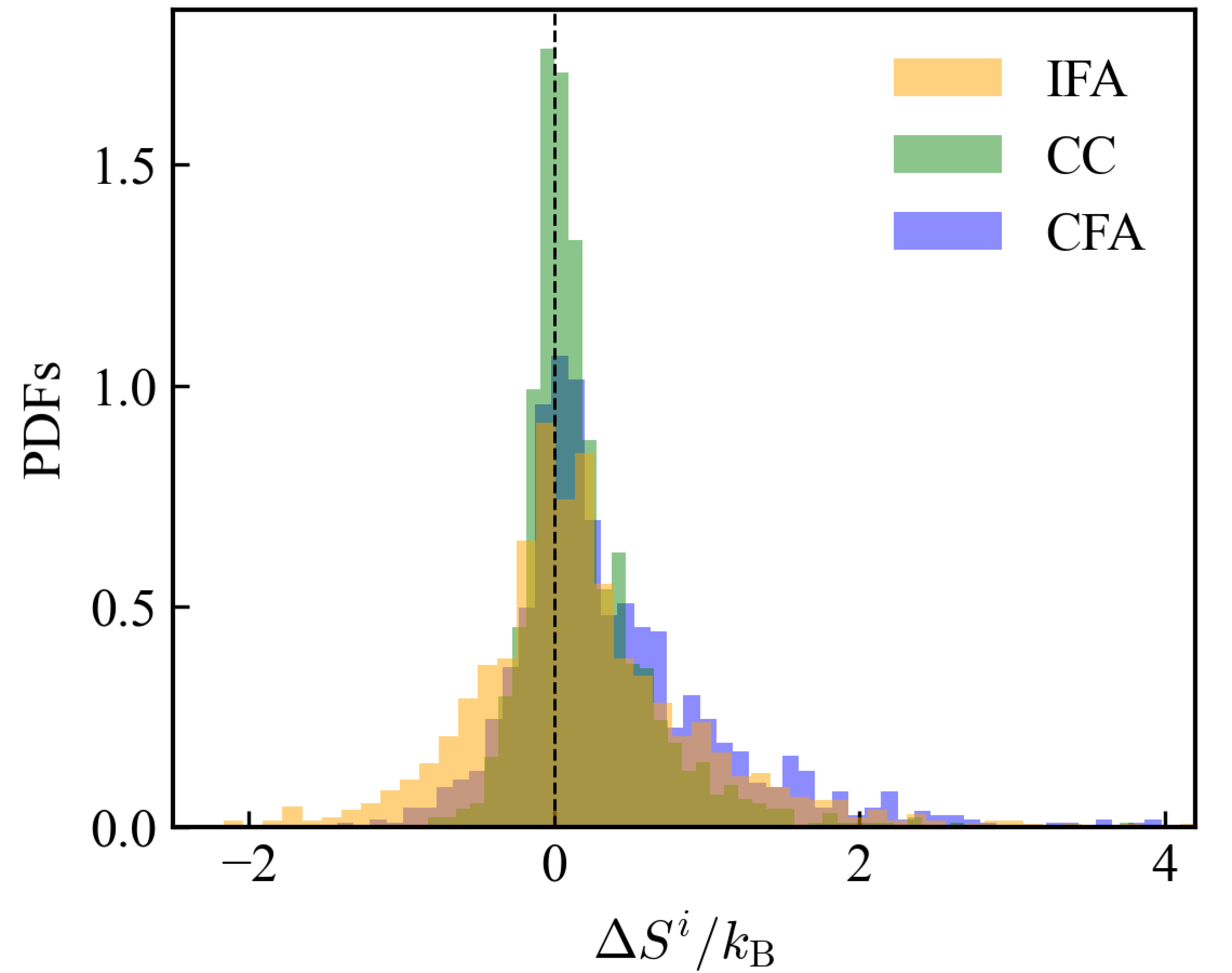}
\caption{
The probability density functions (PDFs) of $\Delta S^i$ for particle $i$ movement toward respective centers in a monodisperse system $(N, \nu)=(1024, 0.740)$.
}
\label{fig:ent}
\end{center}
\end{figure}

We next examine the entropic changes when the particle $i$ is moved to respective centers. The free volume, proportional to the microstate number, is closely related to the entropy~\cite{matsuoka_1997}. The entropic contribution $\Delta S^i$ is defined as:
\begin{equation}
\Delta S^i \equiv k_{\mathrm{B}} \ln \prod_{j=1}^N \frac{{v_{\mathrm{f}}^j}'}{v_{\mathrm{f}}^j},
\end{equation}
where $k_{\mathrm{B}}$ is the Boltzmann constant and $v_{\mathrm{f}}^j$ and ${v_{\mathrm{f}}^j}'$ are free areas before and after $i$'s movement. While the $i$'s free area remains constant, the free area of its neighbors changes, altering the entropy of the system.
Figure~\ref{fig:ent} shows the probability density functions (PDFs) of $\Delta S^i$ for all 1024 particles. Some particles exhibit negative entropy changes for all center types, indicating that no center uniquely increases entropy. However, all PDFs show positive bias, suggesting that overall entropy increases when particles move to their centers. CFA demonstrates the greatest positive bias, followed by CC and IFA. The percentages of particles with $\Delta S^i \geq 0$ are $70.8\%$ for CFA, $65.4\%$ for CC, and $57.5\%$ for IFA, respectively.

\section{Relaxation (Recursive) Dynamics}
\label{sec:3-0}
Previous studies have observed steepest descent dynamics through tracking of energy or root mean square velocity~\cite{chacko_2019, folena_2020, nishikawa_2022, manacorda_2022}. Slow relaxation exhibiting nontrivial power-law decay has been reported in various systems, including dense athermal suspensions of repulsive soft particles, spin-glass models, random Lorentz gas in infinite dimensions, and glass-forming models at high temperatures.
Eq.~(\ref{eqn:rec}) can be rewritten as:
\begin{equation}
\mathbf{u}^i_k = (\tilde{\mathbf{r}}^i_{k+1} - \tilde{\mathbf{r}}^i_k)/\alpha,
\end{equation}
where the displacement direction vector $\mathbf{u}^i_k$ represents the change in position $\tilde{\mathbf{r}}^i_k$ with respect to $\alpha$. We treat $\mathbf{u}^i_k$ as a velocity-like quantity and analyze the step evolution of its mean magnitude $\langle u \rangle^\ast$ with respect to $k\alpha$. $\langle u \rangle^\ast$ generally agrees with the root mean square of $\mathbf{u}^i_k$, exhibiting similar behavior but with reduced fluctuations. Consequently, we observe the relaxation dynamics of recursive algorithms using $\langle u \rangle^\ast$ for the respective centers.

We investigate the relaxation dynamics from initial equilibrated states in both monodisperse and bidisperse hard disk systems. Equilibrated configurations were generated using extended event-driven molecular dynamics (EDMD) simulations~\cite{alder_1959, isobe_1999}. The basic units of the system are the mass of a single disk $m$, the diameter of the disk $2\sigma$, and the energy $1/\beta$. From these, we derive the unit of time $t$ as $2\sigma\sqrt{\beta m}$, where $\beta=1/(k_{\mathrm{B}}T)$. In this study, we set $1/\beta=1$ and $m=1$, giving time $t$ as the dimension of length.

We studied both mono and binary hard disk systems. In the monodisperse system, $1024$ disks with a radius $\sigma$ were placed in a rectangular box $L_x \times L_y$ ($L_y/L_x = \sqrt{3}/2$) with periodic boundary conditions. The packing fraction was set to $\nu= N \pi \sigma^2 /(L_x L_y)=0.740$, which corresponds to the solid phase.
For the bidisperse system, we considered 4096 additive binary hard disks with mole fractions $x_0 = 2/3$ (small) and $x_1 = 1/3$ (large) in a square box $L_x \times L_y$ ($L_y/L_x = 1$) with periodic boundaries. The packing fraction was $\nu= N \pi (x_0\sigma_0^2+x_1\sigma_1^2)/(L_x L_y)=0.790$, with the size ratio $\sigma_1/\sigma_0 = 1.4$, where $\sigma_0$ and $\sigma_1$ are the radii of small and large disks, respectively.
Under these supercompressed conditions, the bidisperse system exhibits characteristics of an amorphous supercooled liquid state in equilibrium~\cite{isobe_2016a, isobe_2016b}. For this system, we used an effective radius $\sigma_{\mathrm{e}}$ to non-dimensionalize $\langle u\rangle^\ast$ in Eq.~(\ref{eqn:condition}). $\sigma_{\mathrm{e}}$ was estimated by averaging $(\sigma_i+\sigma_j)/2$ for collision partners $(i,j)$ through EDMD, following Ref.~\onlinecite{isobe_2016b}. For code optimization, we implemented a grid mapping technique known as the exclusive grid particle method~\cite{isobe_1999} to construct neighbor lists for each particle.

\subsection{Monodisperse systems}
\begin{figure*}
\centering
\includegraphics[scale=0.1]{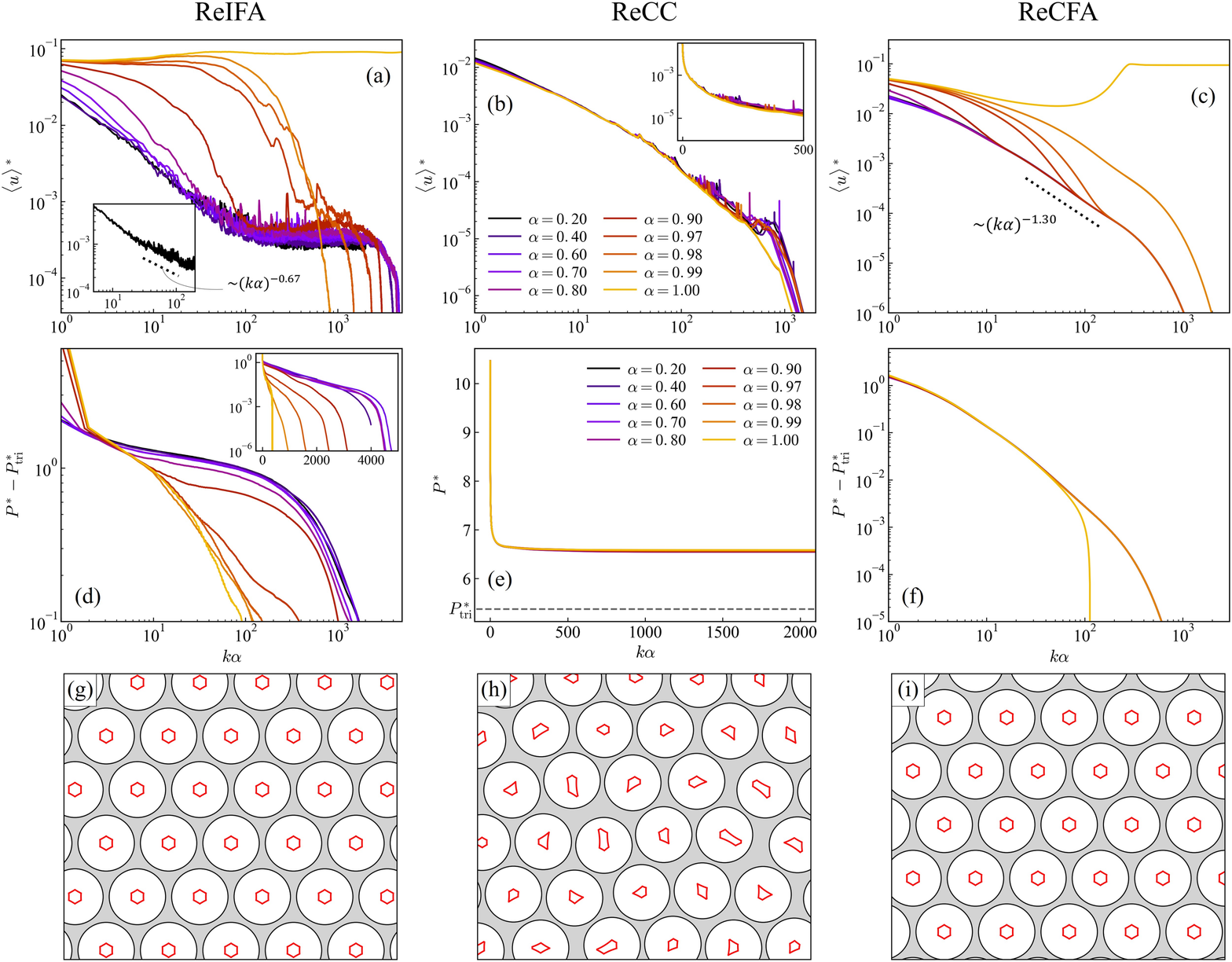}
\caption{
$\alpha$-dependencies of (a-c) dimensionless average magnitude of displacement direction vectors $\langle u\rangle^\ast$ and (d-f) dimensionless static pressure $P^\ast$ as functions of $k\alpha$ in a monodisperse solid phase $(N, \nu)=(1024, 0.740)$ for (a, d) ReIFA, (b, e) ReCC, and (c, f) ReCFA. For ReIFA and ReCFA, the relaxation of $P^\ast$ in (d, f) is plotted on a logarithmic scale after subtracting the triangular-state pressure $P^\ast_{\mathrm{tri}}$. Insets show: (a) extracted result for $\alpha=0.20$, (b, d) semilogarithmic plots. The color scheme for ReIFA and ReCFA matches that of ReCC. (g-i) Converged particle configurations with $\xi = 10^{-5}$ for (g) ReIFA $(\alpha, K) = (0.99, 912)$, (h) ReCC $(\alpha, K) = (1.00, 535)$, and (i) ReCFA $(\alpha, K) = (0.98, 571)$.
}
\label{fig:mono}
\end{figure*}
Figure~\ref{fig:mono} presents the results for the monodisperse system. 
In the monodisperse system, to comprehensively analyze the relaxation dynamics from a thermodynamic perspective, we also investigated
the step evolution of the dimensionless static pressure $P^\ast$, calculated using the free volume~\cite{hoover_1972, speedy_1980, corti_1999, mugita_2024}:
\begin{equation}
P^\ast=\frac{4\nu}{\pi}\left(1+\frac{\sigma}{2N}\sum_{i=1}^{N}\frac{s^i_{\mathrm{f}}}{v^i_{\mathrm{f}}}\right).
\label{eqn:PFV}
\end{equation}
Relaxation of $\langle u \rangle^\ast$ did not show a significant dependence on $\alpha$ in ReCC, while ReIFA and ReCFA exhibited dependence on $\alpha$. For $\alpha = 1.00$, the positions did not converge in both algorithms due to vibrational modes. For $\alpha \neq 1.00$, both algorithms eventually converged to a perfect crystalline structure (triangular lattice) with $\xi=10^{-5}$ (Fig.~\ref{fig:mono}(g) and (i)).

In ReIFA, early stage relaxation showed decay in the power law in $\langle u \rangle^\ast$ for $\alpha \lesssim 0.60$, with an exponent of $-0.67$ for $\alpha = 0.20$ (Fig.~\ref{fig:mono}(a) and its inset). Mid-stage relaxation exhibited a plateau, with higher $\alpha$ values leading to faster escape and rapid exponential decay. $\alpha = 0.99$ showed the fastest convergence without a plateau. 
$P^\ast$ relaxation towards $P^\ast_{\mathrm{tri}}$ exhibited a power-law decay for small $\alpha$ ($k\alpha \lesssim 10^2$), corresponding to $\langle u \rangle^\ast$ behavior. Two-stage exponential relaxation was observed in pressure (Fig.~\ref{fig:mono}(d) and its inset).

ReCFA showed similar relaxation dynamics for $\alpha \lesssim 0.70$, with power-law decay (exponent $-1.30$) in mid-stage relaxation (Fig.~\ref{fig:mono}(c)). The larger $\alpha$ values increased the time to reach power-law decay. At $\alpha = 0.99$, a distinct final exponential decay occurred without reaching other power-law decays.
$P^\ast - P^\ast_{\mathrm{tri}}$ exhibited consistent relaxation curves for all $\alpha$ except 1.00 (Fig.~\ref{fig:mono}(f)).

ReCC showed similar dynamics for all $\alpha$ values, without clear power-law decay in $\langle u \rangle^\ast$ relaxation (Fig.~\ref{fig:mono}(b)). The relaxation was not purely exponential (Fig.~\ref{fig:mono}(b) inset). The state does not reach the perfect crystalline structure (Fig.~\ref{fig:mono}(e) and (h)).

\subsection{Bidisperse systems}
\begin{figure*}
\centering
\includegraphics[scale=0.1]{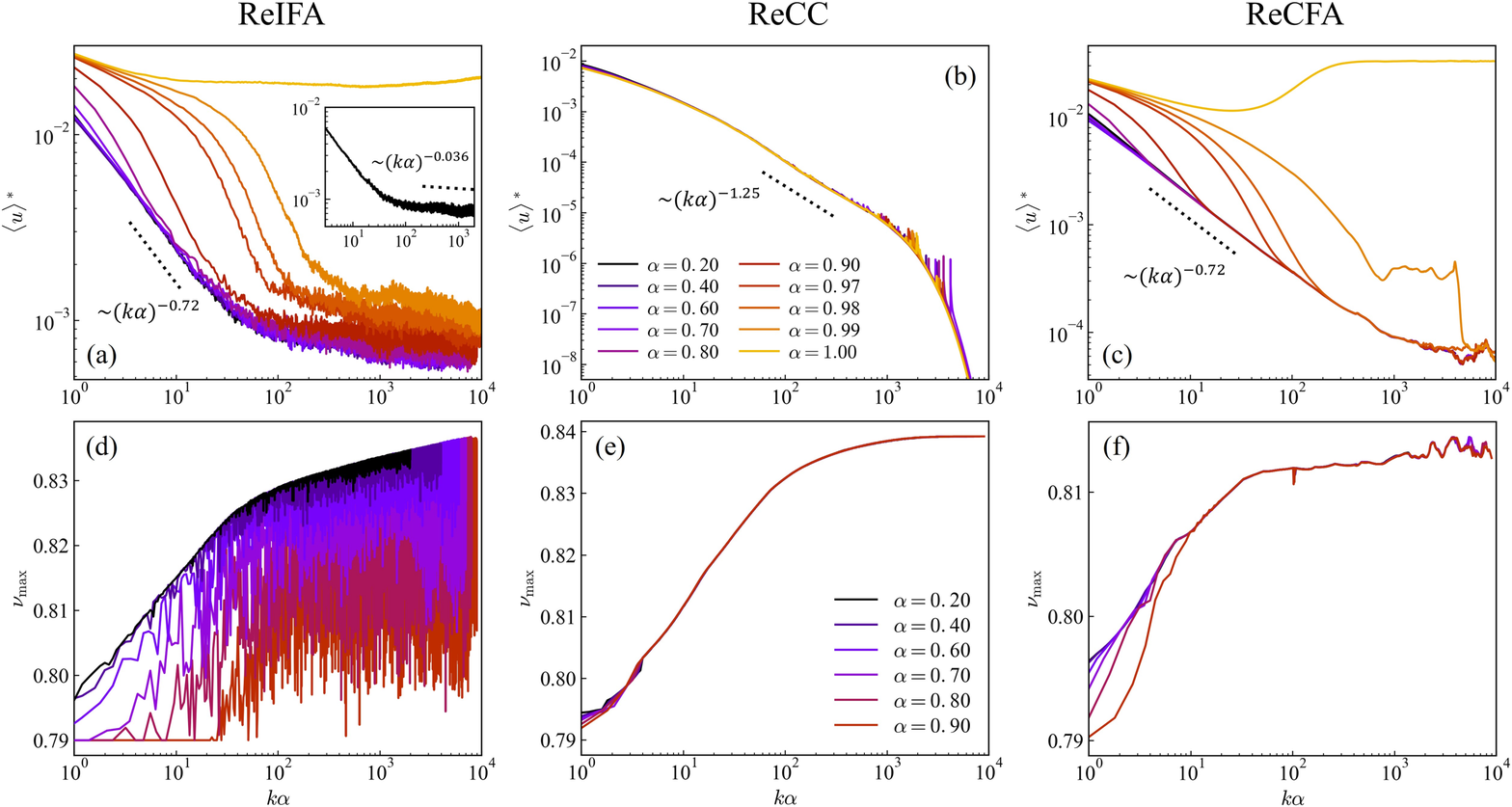}
\caption{
$\alpha$-dependencies in a bidisperse system $(N, \nu)=(4096, 0.790)$ as functions of $k\alpha$ for: (a-c) the dimensionless average magnitude of displacement direction vectors $\langle u\rangle^\ast$ by (a) ReIFA, (b) ReCC, and (c) ReCFA, and (d-f) the maximum packing fraction limit $\nu_{\mathrm{max}}$ after uniform expansion of particle diameter by (d) ReIFA, (e) ReCC, and (f) ReCFA. Inset in (a): extracted result for $\alpha=0.20$. The results represent averaged relaxation behavior from five independent initial equilibrium states. The color scheme for ReIFA and ReCFA matches that of ReCC.
}
\label{fig:bin}
\end{figure*}
Figure~\ref{fig:bin} presents results for bidisperse systems, averaged over five independent initial equilibrium configurations. 
In the bidisperse system, 
we
examined
not only 
the relaxation of $\langle u \rangle^\ast$, but also the relaxation of
$\nu_{\mathrm{max}}$ 
to gain a better understanding of the properties of nearly jammed packings generated by each method, where $\nu_{\mathrm{max}}$ is the packing fraction when all particles are uniformly expanded to their maximum limit without overlap at each step.

\begin{figure}
\begin{center}
\includegraphics[scale=0.145]{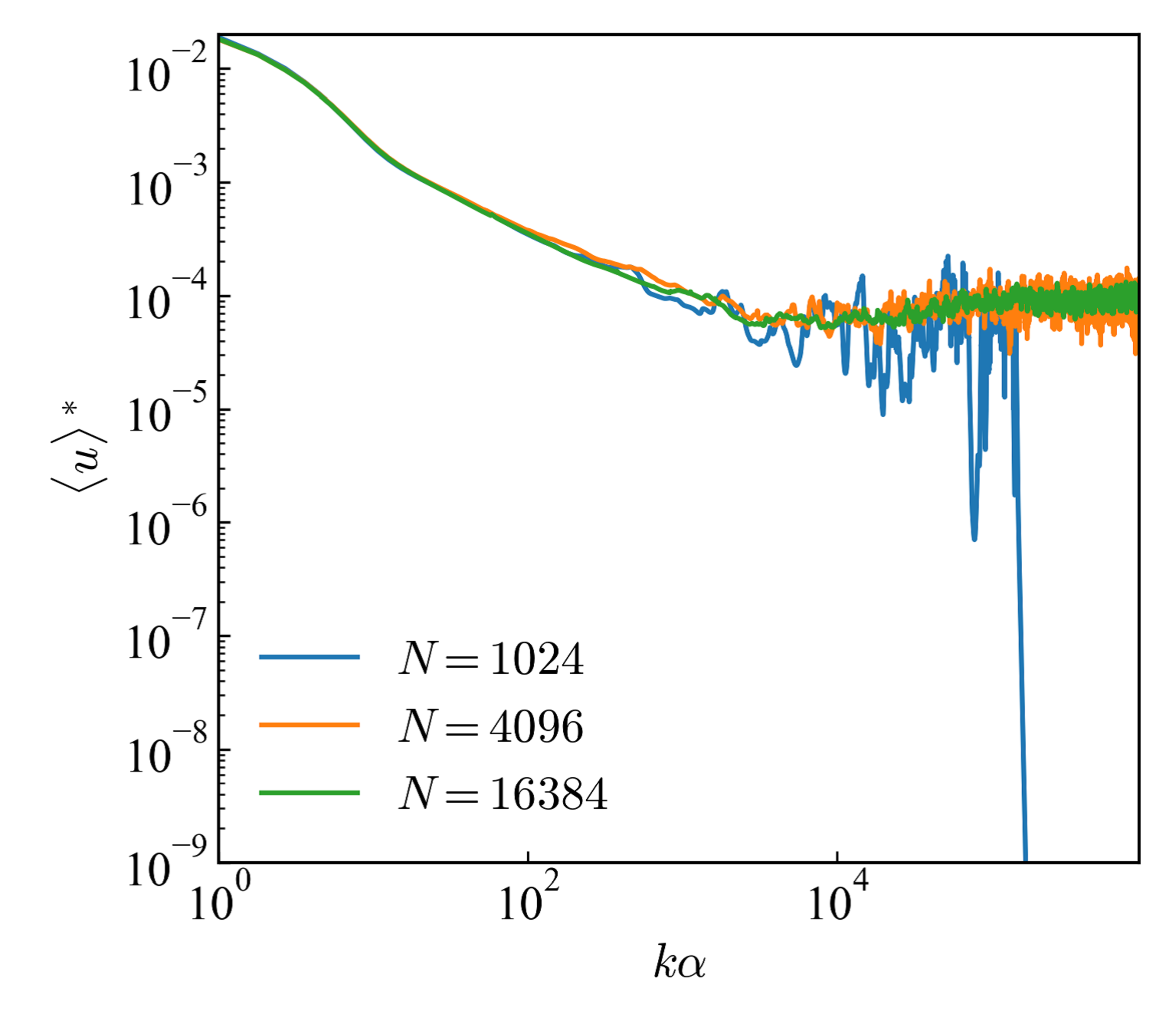}
\caption{
System size dependence of $\langle u\rangle^\ast$ in ReCFA at $\alpha=0.90$.
}
\label{fig:system_size}
\end{center}
\end{figure}

Similar to the relaxation of $\langle u \rangle^\ast$ in the monodisperse system, ReCC did not show a dependence on $\alpha$, while ReIFA and ReCFA exhibited a dependence on $\alpha$ (Fig.~\ref{fig:bin}(a-c)). For $\alpha=1.00$ in ReIFA and ReCFA, the system did not converge due to vibrational modes. However, other relaxation properties differed from the monodisperse system.
In ReIFA, after an initial power law decay (exponent $-0.72$), the system transitioned to another power law decay with larger fluctuations rather than a complete plateau. For $\alpha = 0.20$, the second exponent was $-3.6\times10^{-2}$ (Fig.~\ref{fig:bin}(a) inset). ReCFA exhibited an initial power-law decay exponent equivalent to that of ReIFA at $-0.72$. After this decay, the system entered a regime of stationary states instead of transitioning to exponential decay. In contrast, ReCC showed a clear decay of the power law with an exponent of $-1.25$.
For $k\alpha \le 10^4$, neither ReIFA nor ReCFA reached exponential decay. We investigated the system size dependence of longer relaxation dynamics using ReCFA for $\alpha=0.90$ (Fig.~\ref{fig:system_size}). For $N=1024$, the system escaped stationary states after power-law decay and transitioned to rapid exponential decay. This suggests that sufficiently long calculations for $N=4096$ (and $16384$) would eventually lead to exponential decay in both ReCFA and ReIFA.
Figure~\ref{fig:system_size} shows no dependence on system size on the onset of stationary states or $\langle u \rangle^\ast$ values. The stationary behavior after power-law decay increased slightly rather than forming a complete plateau.

Examining the results of $\nu_{\mathrm{max}}$, all recursive algorithms exhibited (logarithmic) growth with respect to $k\alpha$ in the early stage (Fig.~\ref{fig:bin}(d-f)), indicating the generation of nearly jammed packing. ReIFA and ReCFA demonstrated a two-stage relaxation regime: an initial rapid logarithmic increase followed by a slower logarithmic increase. Based on the relaxation analysis $\langle u \rangle^\ast$, we infer that neither system has fully converged, and continued calculations would likely lead to further increases in $\nu_{\mathrm{max}}$.
In ReIFA, $\nu_{\mathrm{max}}$ fluctuations increased significantly with increasing $\alpha$, suggesting that this method becomes unstable and unsuitable to generate jammed packings at relatively large values $\alpha$.

\begin{figure}
\begin{center}
\includegraphics[scale=0.16]{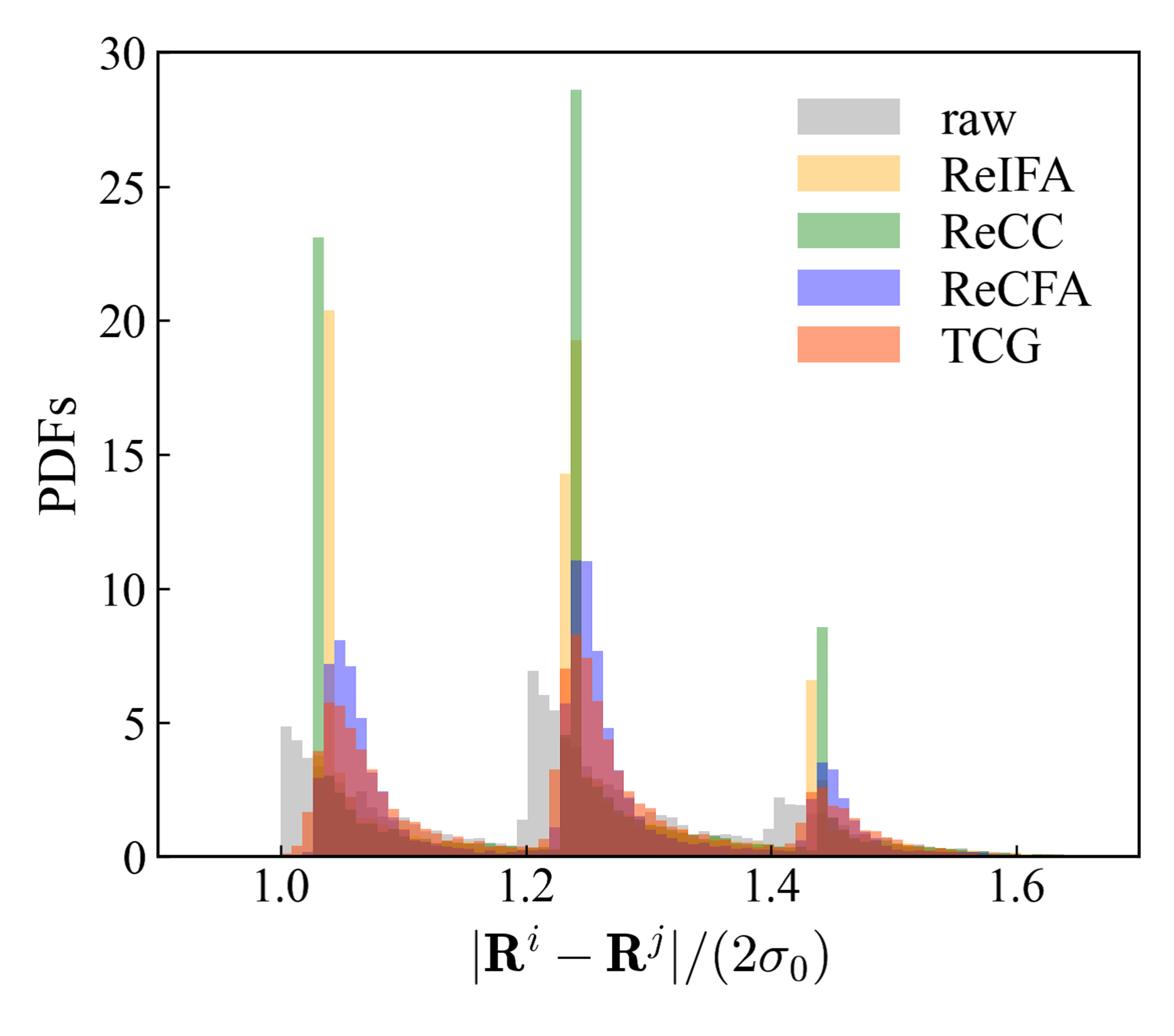}
\caption{
PDFs of neighbor distances are analyzed for three configurations: raw ($\mathbf{r}^i$), TCG ($\overline{\mathbf{r}^i}$), and those obtained from recursive algorithms ($\mathbf{r}^i_{\mathrm{jam}}$), where $\mathbf{R}^i$ denotes any of these configurations.
}
\label{fig:PDF}
\end{center}
\end{figure}
To investigate the properties of nearly jammed packings, we calculated the PDF of nearest neighbor distances for configurations generated by three algorithms (Fig.~\ref{fig:PDF}). Neighbors were identified using 2D-SANNex~\cite{mugita_2024}. The convergence thresholds $\xi$ were set at $10^{-5}$ for ReCC, $10^{-3}$ for ReIFA, and $10^{-4}$ for ReCFA, with optimal parameter sets $(\alpha, K)$ of $(0.70, 68)$, $(1.00, 694)$ and $(0.98, 789)$, respectively.
Three peaks corresponding to small-small, small-large, and large-large particle pairs were observed. All recursive algorithms produced sharper peaks compared to the initial equilibrium configuration, with ReCC showing the sharpest peaks, followed by ReIFA and ReCFA. This trend correlates with the final values of $\nu_{\mathrm{max}}$. We conclude that ReCC is the most advantageous for generating jammed packings due to its fastest and most stable convergence, as well as the highest $\nu_{\mathrm{max}}$.

For the reference of comparison, we applied TCG to the particle configuration:
\begin{equation}
\overline{\mathbf{r}^i}(t) = \frac{1}{\delta t}\int^{\delta t/2}_{-\delta t/2} \mathbf{r}^i(t+t') dt',
\end{equation}
with $\delta t = 100 \times t_{\mathrm{MFT}}$ ($\delta t/(2\sigma_{\mathrm{e}})=1.85$ at $\nu=0.790$), where $t_{\mathrm{MFT}}$ is the mean free time. The TCG configuration showed sharper peaks than the raw configuration, but broader than those from ReCFA.
For the TCG configuration, attempts to increase $\nu$ by uniformly expanding all particles in increments of $10^{-4}$ in 10 independent configurations failed to produce any increase beyond $\nu = 0.790$.

\section{Hopping Motion Analysis}
\label{sec:4-0}
\begin{figure*}
\centering
\includegraphics[scale=0.115]{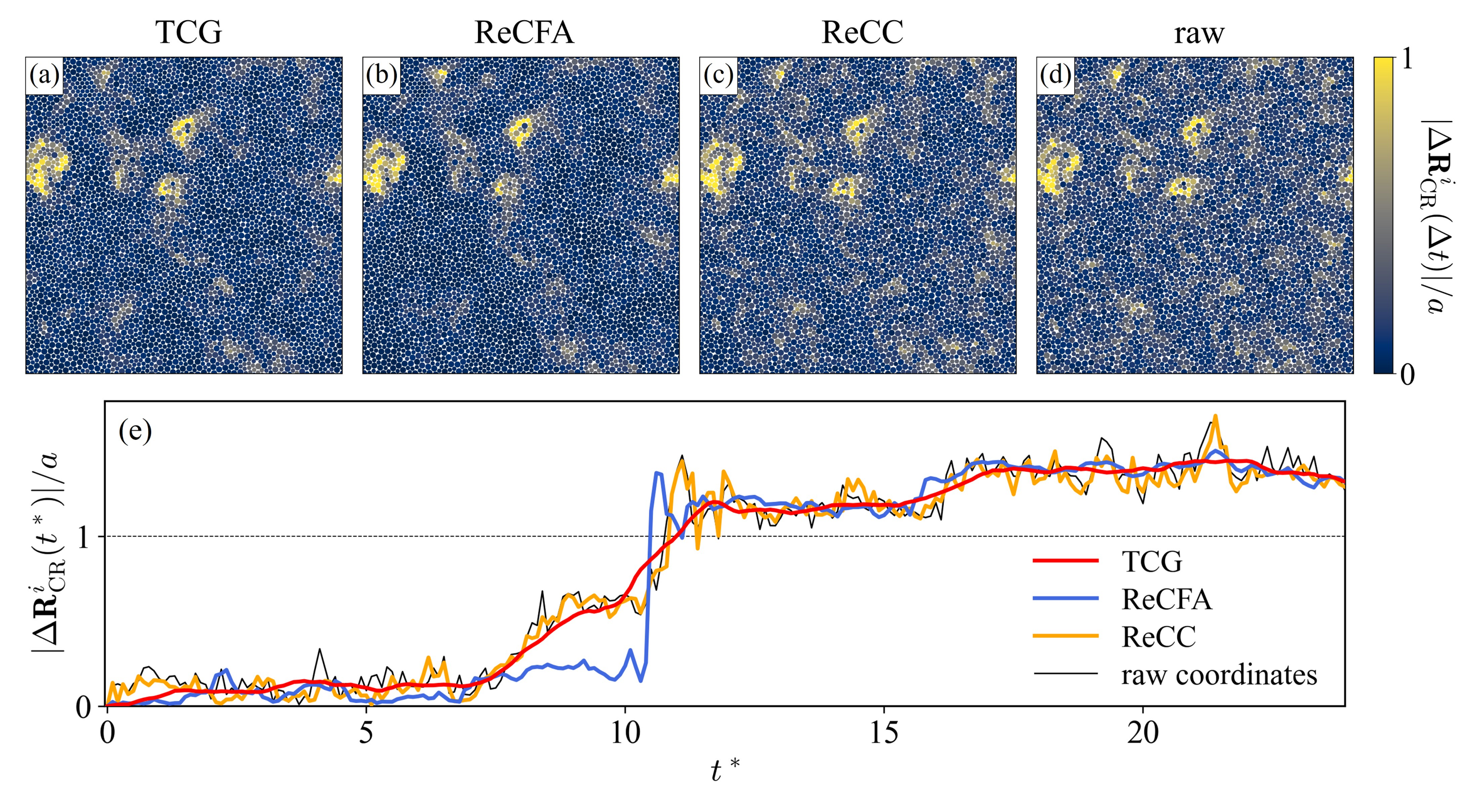}
\caption{Spatial cage-relative displacement field $|\Delta \mathbf{R}^i_{\mathrm{CR}}(\Delta t)|/a$ in (a) TCG, (b) ReCFA, (c) ReCC, and (d) raw coordinates, where yellow (bright) particles indicate larger displacements. (e) Typical temporal evolution of cage-relative displacement for a representative particle $i$, derived using methods (a-d), where $t^* = t/(2\sigma_{\mathrm{e}})$.
}
\label{fig:spatial_dis}
\end{figure*}

\begin{figure*}
\centering
\includegraphics[scale=0.15]{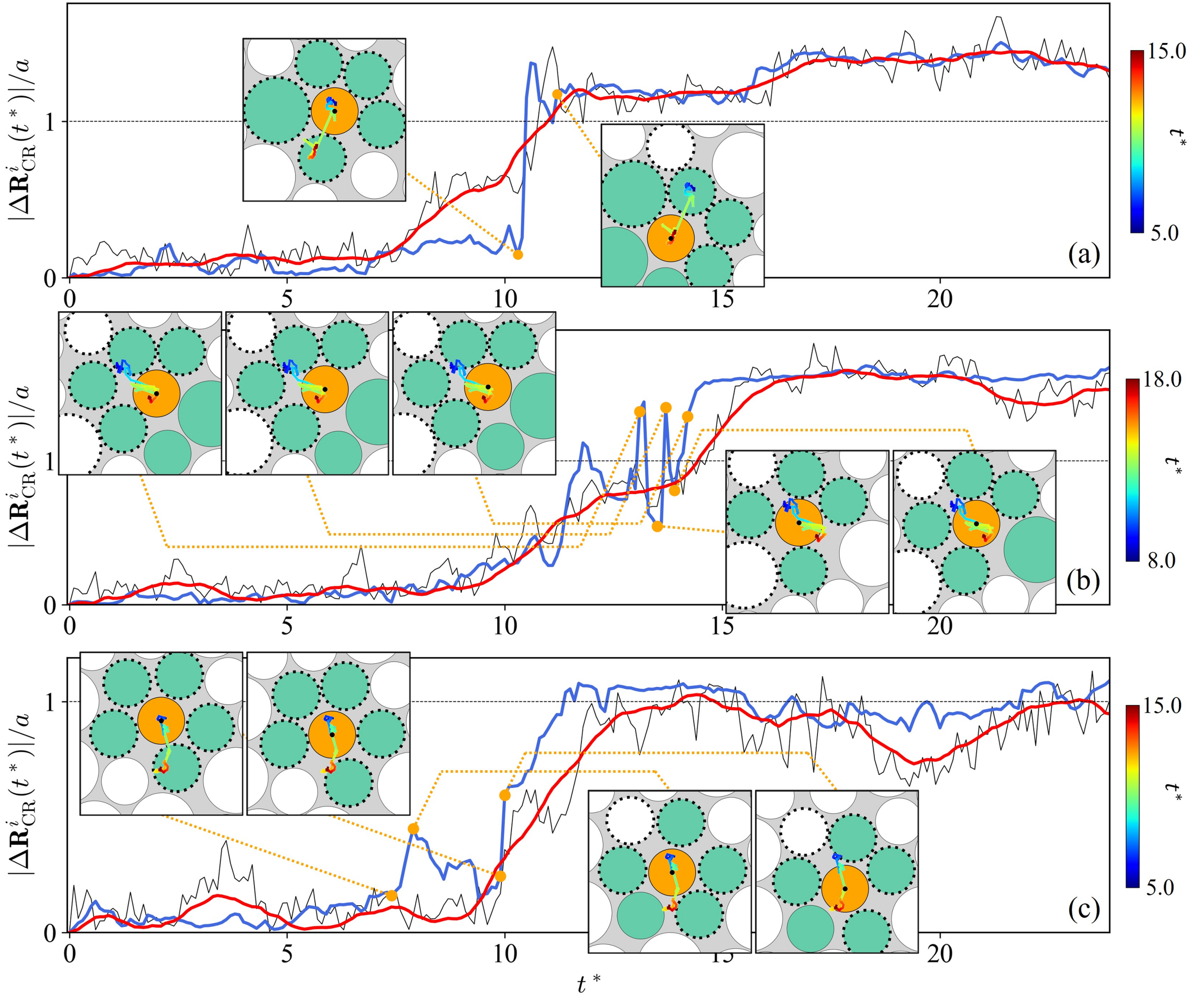}
\caption{
Typical cage-relative displacements with hop detection by TCG and ReCFA, where (a) is identical to Fig.~\ref{fig:spatial_dis}. The orange (center) particle denotes the target particle $i$; the green particles indicate its instantaneous nearest neighbors; the black dotted circles represent initial nearest neighbors at $t^\ast=0$. ReCFA trajectories are color-coded according to temporal evolution.
}
\label{fig:typical}
\end{figure*}

We compared string-like hopping motions in trajectories obtained by TCG and recursive algorithms in a bidisperse equilibrium state under supercompressed conditions. ReCFA parameters were adjusted to $\xi = 3 \times 10^{-4}$ and $(\alpha, K) = (0.97, 137)$ to be consistent with the TCG displacement relative to the raw configuration, while the ReIFA and ReCC parameters remained unchanged from the previous section.
Hopping motion detection utilized the single-particle indicator function~\cite{keys_2011, speck_2012}:
\begin{equation}
h^i=\theta[|\Delta \mathbf{R}^i_{\mathrm{CR}}(\Delta t)|-a],
\end{equation}
where $\theta$ is the Heaviside step function. To eliminate Mermin-Wagner fluctuations~\cite{mermin_1966, mermin_1968}, we employed cage-relative displacements~\cite{mazoyer_2010, shiba_2018, flenner_2019}:
\begin{equation}
\Delta \mathbf{R}^i_{\mathrm{CR}}(t)=[\mathbf{R}^i(t)-\mathbf{R}^i(0)]-\frac{1}{N^i_{\mathrm{NN}}}\sum_{j=1}^{N^i_{\mathrm{NN}}}[\mathbf{R}^j(t)-\mathbf{R}^j(0)],
\end{equation}
where $N^i_{\mathrm{NN}}$ is the number of nearest neighbors of particle $i$ at $t=0$, and $\mathbf{R}^i$ represents raw positions $\mathbf{r}^i$, TCG positions $\overline{\mathbf{r}^i}$, or nearly jammed packing positions $\mathbf{r}^i_{\mathrm{jam}}$. We set $(a/(2\sigma_{\mathrm{e}}), \Delta t/(2\sigma_{\mathrm{e}}))=(0.60, 41.0)$.
Figure~\ref{fig:spatial_dis}(a-d) compares spatial distributions of hopping motions across various coordinate systems. ReCC results show particle displacements similar to the raw coordinates, indicating incomplete removal of thermal fluctuations. In contrast, ReCFA, like TCG, eliminates small displacements, providing clearer hopping motion analysis. Figure~\ref{fig:spatial_dis}(e) demonstrates that while ReCC closely traced the raw coordinates, ReCFA successfully removed small vibrations, similar to TCG.
Hopping in ReCFA coordinates ends more rapidly than in raw coordinates, suggesting successful configuration space tiling. This characteristic, also observed in the results of the inherent structure in soft particle systems~\cite{keys_2011}, is confirmed by examining the particle configurations before and after hopping (Fig.~\ref{fig:typical}(a) and Supplemental Material~\cite{SM}).
High-frequency, large-amplitude oscillations during hopping in ReCFA coordinates, as shown in Fig.~\ref{fig:typical}(b), indicate a clear demarcation between adjacent tiles, suggesting effective tiling. Although generally undesirable in hopping analysis, these fluctuations provide insight into the configuration space structure.
Interestingly, oscillations in ReCFA often precede apparent hopping onset in raw or TCG coordinates, suggesting prediction of post-hop structures. 
This phenomenon was observed in several other hopping cases, as illustrated in Fig.~\ref{fig:typical}(c), where particle configuration changes are expected before the actual hop occurs.


\section{Concluding Remarks}
\label{sec:5-0}
In this study, we introduce and apply a novel recursive algorithm called ``Recursive algorithm to the Centroid of Free Area (ReCFA).'' Inspired by the steepest descent method, ReCFA is designed to derive the inherent structure of configurational space in free-energy landscapes with flat potential energy and to analyze hopping motions in highly packed mono and binary hard disk systems. We compare the characteristics of ReCFA with the conventional time-coarse-graining (TCG) method and alternative recursive algorithms using the ``inner center'' of the free area (ReIFA) and cage center (ReCC). Our results demonstrate that ReCFA exhibits several advantages in entropic contribution to the minimization of free energy and in clearly detecting hopping motions in supercompressed binary hard disk systems.

Comparative analysis of recursive algorithms---ReCFA, ReIFA, and ReCC---reveals distinct differences in their convergence and final particle configurations. From the perspective of generating jammed packing states, ReCC demonstrates superior performance among the three algorithms, achieving the highest packed density with fewer iterations. However, it proves unsuitable for detecting hopping motion due to its sensitivity to small perturbations in particle displacements. In contrast, ReCFA demonstrates high efficacy in successfully removing thermal fluctuations from instantaneous configurations.
Furthermore, we find that the particle coordinates derived by ReCFA do not necessarily follow the raw and TCG coordinates with and without thermal fluctuations. Instead, they exhibit instantaneous displacements within a short time as precursors of large structural changes (changes in nearest neighbors), followed by hopping motion. This indicates that ReCFA correctly detects effective basins of local maximum entropy for each instantaneous time. Notably, some hopping motions detected by ReCFA are not detected by TCG, and vice versa. We aim to elucidate this discrepancy in detail using a more precise method involving the iso-configuration ensemble at higher density.
The relaxation dynamics exhibits several similarities to the dynamics of steepest descent, characterized by slow power-law relaxation followed by rapid exponential relaxation. This observation provides valuable insights for further theoretical investigation and could enhance our understanding of structural relaxation and dynamic heterogeneity in more complex free energy landscapes.

In distinguishing ReCFA from similar algorithms, such as the Lloyd algorithm~\cite{lloyd_1982} (a recursive algorithm based on Voronoi centers), where the nearest neighbors detected reciprocally by the Voronoi method do not correspond to organized nonreciprocal particles of free volume, ReCFA shows promise as an effective method for studying hopping motion and jammed states. Further investigation of these applications will be pursued in future work.

\section*{Acknowledgements}
The authors express their gratitude to Professors C.-H. Lam, C.-T. Yip, T. Kawasaki, Y. Nishikawa, M. Engel, and W. Krauth for their stimulating and fruitful discussions on this topic. M.I. acknowledges support from JSPS KAKENHI Grant Nos. 20K03785 and 23K03246. Computational resources were partially provided by the Supercomputer Center, Institute for Solid State Physics (ISSP), University of Tokyo. This research was conducted as part of the International Research Project on Non-Reversible Markov Chains, Implementations and Applications.

\bibliography{bibtex}

\end{document}